\documentclass[10pt, conference, letterpaper]{IEEEtran}

\usepackage[cmex10]{amsmath}
\usepackage{amsfonts,amssymb,fancyhdr,layout}
\usepackage{url,xspace,graphicx}
\usepackage{color}
\usepackage{comment}
\usepackage{float}
\usepackage{subcaption}
\usepackage{array}
\usepackage{multirow}
\usepackage{pslatex}
\usepackage{clrscode}
\usepackage{amsthm}
\usepackage[hidelinks]{hyperref}
\usepackage{capt-of}

\newcommand{\timef}{\proc{TimeFlip}}
\newcommand{\rptp}{\proc{ReversePTP}}
\newcommand{\timeb}{time-based}
\newcommand{\timec}{\proc{Time4}}

\newcommand{\offs}{\mathsf{offset}}
\newcommand{\offseti}{\mathrm{\offs}_{i}}

\newcommand{\onec}{OneClock}

\newlength{\grafflecm}
\begin{document}

\title{Timing in Software-Defined and \\ Centrally-Managed Networks}

\author{
{Tal Mizrahi, Yoram Moses}\\
Technion --- Israel Institute of Technology
}

\maketitle
\thispagestyle{empty}

\begin{abstract}
The work described in this paper explores the use of time and synchronized clocks in centrally-managed and Software Defined Networks (SDNs). One of the main goals of this work is to analyze use cases in which explicit use of time is beneficial. Both theoretical and practical aspects of timed coordination and synchronized clocks in centralized environments are analyzed. Some of the products of this work are already incorporated in the OpenFlow specification, and open source prototypes of the main components are publicly available.
\end{abstract}

\enlargethispage{-6.4cm}
\noindent\begin{picture}(0,0)
\put(0,-390){\begin{minipage}{\textwidth}
\centering
\fbox{\includegraphics[width=.99\textwidth]{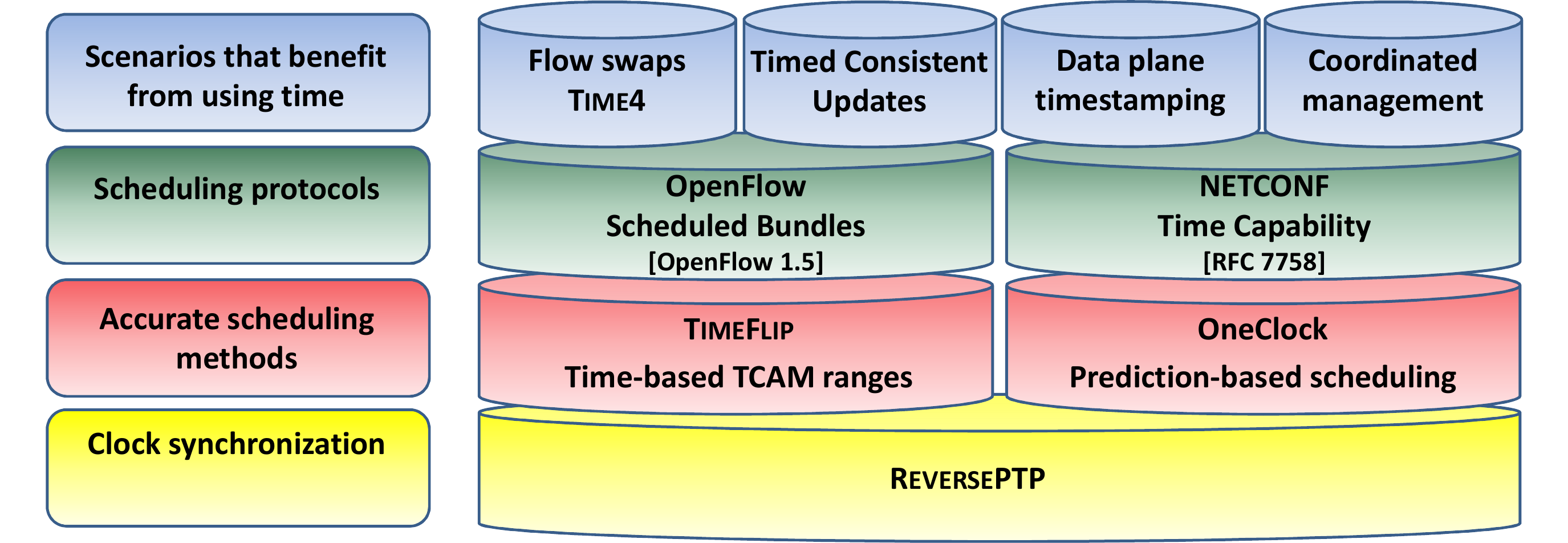}}
\captionof{figure}{The TimedSDN project at a glance.}
\label{fig:Results}
\end{minipage}}
\end{picture}%

\section{Introduction}
\label{IntroSec}

\subsection{Background}
Network time synchronization has evolved significantly over the last decade. Indeed, the Precision Time Protocol (PTP), defined in the IEEE 1588 standard, can synchronize clocks to a very high degree of accuracy, typically on the order of 1 microseconds. Since its publication in 2008, PTP has matured and has become a common and affordable feature in commodity switches. We argue that since most of the SDN products already have built-in hardware capabilities for accurate clock synchronization, it is only natural to harness this powerful technology to coordinate events in SDNs.

This paper presents an overview of the TimedSDN project. The key concept behind this project is simple: accurate time can be used to coordinate network configuration and policy updates in centrally managed environments, and specifically in SDNs. In a nutshell, the SDN controller can invoke time-triggered updates in network switches, allowing near-simultaneous network-wide updates, or allowing the controller to invoke multi-step updates, in which each step is invoked at a different update time.

\subsection{Contributions}

The main contributions of this work are in four main aspects, as depicted in the four rows of Fig.~\ref{fig:Results}.

\enlargethispage{-6.4cm}

\begin{itemize}
	\item \textbf{Scenarios that benefit from using time.} Time is shown to be a powerful and effective tool in coordinated centrally managed networks such as SDNs. The use of synchronized clocks enables near-simultaneous updates, or multi-phase updates, in which multiple nodes are updated at different times. Notably, using time improves the network performance during network updates by reducing the packet loss and resource overhead. 
This key result is presented using theoretic analysis, as well as by experimental evaluation.
	\item \textbf{Scheduling protocols.} The current work defines extensions to standard network protocols, enabling practical implementations of our concepts.
We define a new feature in OpenFlow called Scheduled Bundles, which has been incorporated into the OpenFlow 1.5 protocol. We defined a similar capability for the Network Configuration (NETCONF) protocol, which has been published as an RFC.
  \item \textbf{Accurate scheduling.} A key challenge in time-triggered applications is accurate scheduling, i.e., guaranteeing that events are executed at the exact time for which they were scheduled. In this work we present and analyze two accurate scheduling methods. The first uses Ternary Content Addressable Memory (TCAM) ranges in hardware switches. This method was successfully demonstrated in practice on a real-life device, a Marvell DX switch. The second method is a prediction-based scheduling approach that uses timing information collected at runtime to accurately schedule future operations. Both methods are shown to be practical and efficient.
	\item \textbf{Clock Synchronization.} The current work also introduces \rptp, a clock synchronization scheme that is adapted to centralized environments such as SDNs.
\end{itemize}

These four aspects are discussed in greater detail in Sections~\ref{WhySec}-\ref{SyncSec}.

Open source prototypes of the OpenFlow time extension, the NETCONF time capability, and \rptp\ are publicly available~\cite{TimedSDNSource}.

\subsection{Related Work}
This paper summarizes ``Using Time in Software Defined Networks'', a PhD dissertation~\cite{UsingTimePhD} that was submitted to the Technion in 2016. This work produced several publications~\cite{Time4Journal,TimeFlipToN,TimedConsistentToN,TimedConsistent,Time4Infocom,Infocom-TimeFlip,OneClockNOMS,hotsdn,swfanDPT,rptpijnm,hotsdnrptp,ispcsrptp,TimeCap,TimeTR}. A preliminary version of this overview appeared~\cite{TimedSDNNewsLetter} in the IEEE SDN Newsletter, November 2016.

A network configuration update is \emph{per-packet consistent} \cite{reitblatt2012abstractions} if it guarantees that every packet sent through the network is processed according to a single configuration version, either the previous or the current one. Various approaches to ensuring consistent network updates appear in the literature, e.g.,~\cite{reitblatt2012abstractions,francois2007avoiding, bryant2005ip, jin2014dynamic,vanbever2011seamless,liu2013zupdate}. However, the current work is the first to consider the use of accurate time and synchronized clocks in network updates in general, and specifically in updates that are required to be consistent.

The use of time in distributed applications has been widely analyzed, both in theory and in practice. Analysis of the usage of time and synchronized clocks, e.g., Lamport~\cite{lamport1978time, LamportTimeout} dates back to the late 1970s and early 1980s. In recent years, as accurate time has become an accessible and affordable tool, it is used in various different applications, e.g.,~\cite{corbett2013spanner,harris2008application,IEEETSN,moreira2009white}. While the usage of accurate time in distributed systems has been widely discussed in the literature, we are not aware of similar analyses of the usage of accurate time as a means for performing accurately scheduled configuration updates in computer networks.

Prior to the current work, neither the OpenFlow protocol~\cite{OpenFlow1.4} nor common management and configuration protocols, such as SNMP~\cite{snmp} and NETCONF~\cite{netconf}, used accurate time for scheduling or coordinating configuration updates. The current work defines extensions to the OpenFlow protocol and to the NETCONF protocol which enable the use of time-triggered updates in real-life networks.

\section{Scenarios that Benefit from Using Time}
\label{WhySec}

We briefly present a few key use cases that greatly benefit from using time.

\begin{figure}[htbp]
  \centering
	\fbox{\includegraphics[width=.47\textwidth]{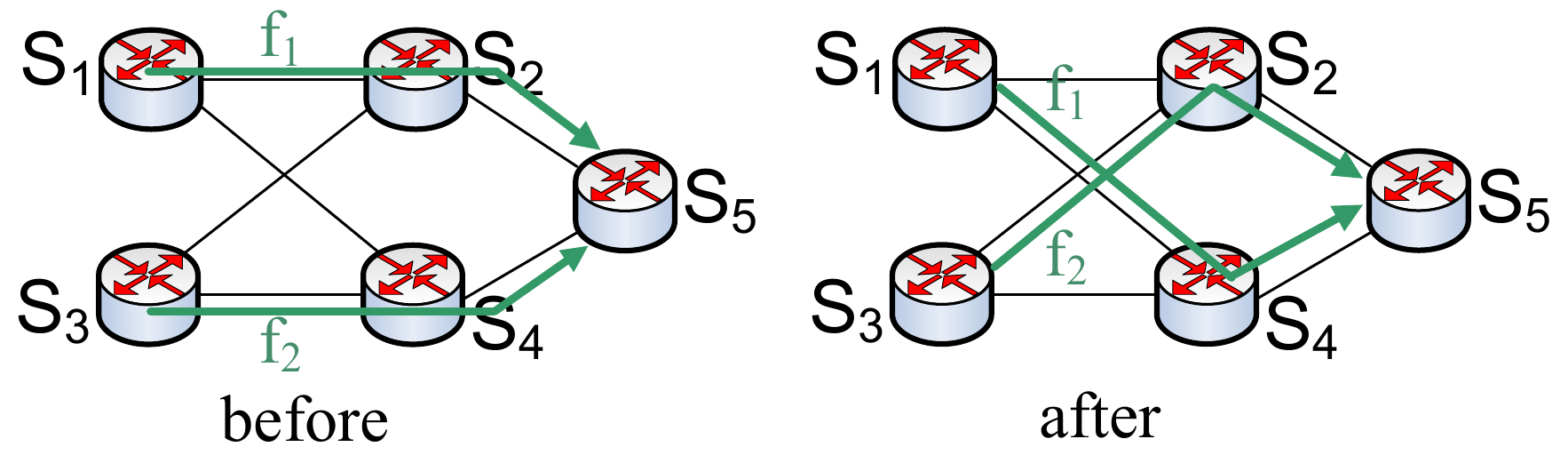}}
	\captionsetup{justification=centering}
  \caption{Flow swapping: flows need to convert from the `before' configuration to the `after' configuration. Updating $S_1$ and $S_2$ at the same time is the optimal approach.}
  \label{fig:Swap}
\end{figure}

\textbf{Flow swaps.} 
\timec~\cite{Time4Infocom,Time4Journal} is a network update approach that performs multiple changes at different switches at the same time.
The work of~\cite{Time4Infocom,Time4Journal} considers a class of network update scenarios called \emph{flow swaps}, and shows that \timec\ is the optimal approach for implementing them; \timec\ allows fewer packet drops and higher scalability than state-of-the-art approaches. A flow swapping example is illustrated in Fig.~\ref{fig:Swap}.

In this work we presented the \emph{lossless flow allocation} (LFA) problem, and used a game-theoretic analysis to formally show that flow swaps are provably inevitable in the dynamic nature of SDN. 


\textbf{Timed consistent updates.} 
A consistent network update~\cite{reitblatt2012abstractions} is an update that guarantees that every packet is forwarded either according to the previous configuration before the update, or according to the new configuration after the update, but not according to a mixture of the two.
The approach we presented in~\cite{TimedConsistent,TimedConsistentToN} introduces \textbf{time-triggered} multi-phase network updates, which can guarantee consistency while requiring a shorter duration than existing consistent update methods.
This approach is formally shown to reduce the expensive overhead of maintaining duplicate configurations.
	

\begin{figure}[htbp]
  \centering
	\fbox{\includegraphics[width=.47\textwidth]{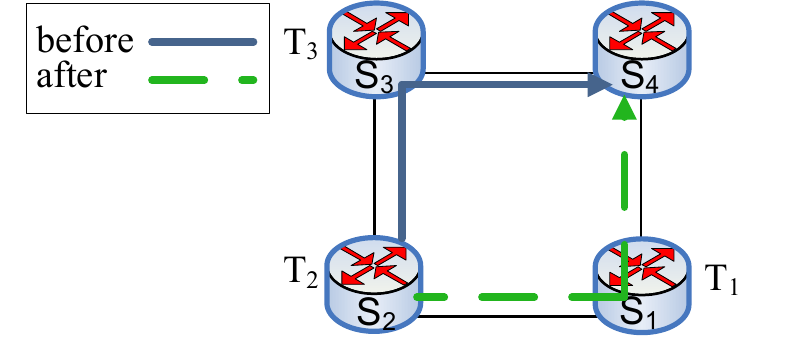}}
	\captionsetup{justification=centering}
  \caption{A timed consistent update example: when updating from the `before' to the `after' configuration, each local update is performed at a different time, $T_1$, $T_2$, and $T_3$.}
  \label{fig:MultiPhase}
\end{figure}

\textbf{Data plane timestamping.}
In~\cite{swfanDPT} we argued that in the unique environment of SDN, attaching a timestamp to the header of all packets is a powerful feature that can be leveraged by various diverse SDN applications. We analyzed three key use cases that demonstrate the advantages of using DPT: (i) network telemetry, (ii) consistent network updates, and (iii) load balancing. We also showed that SDN applications can benefit even from using as little as one bit for the timestamp field.

\textbf{Coordinated management.}
As discussed in~\cite{OneClockNOMS}, time can be a valuable tool in coordinating network management, not only in SDNs, but in any centrally managed network. For example, synchronized time can be used to coordinate a configuration update that should be performed at multiple nodes at the same time, or to take a coordinated snapshot of the state of multiple nodes in the network. 

\begin{figure*}[htbp]
  \centering
	\fbox{\includegraphics[width=.8\textwidth]{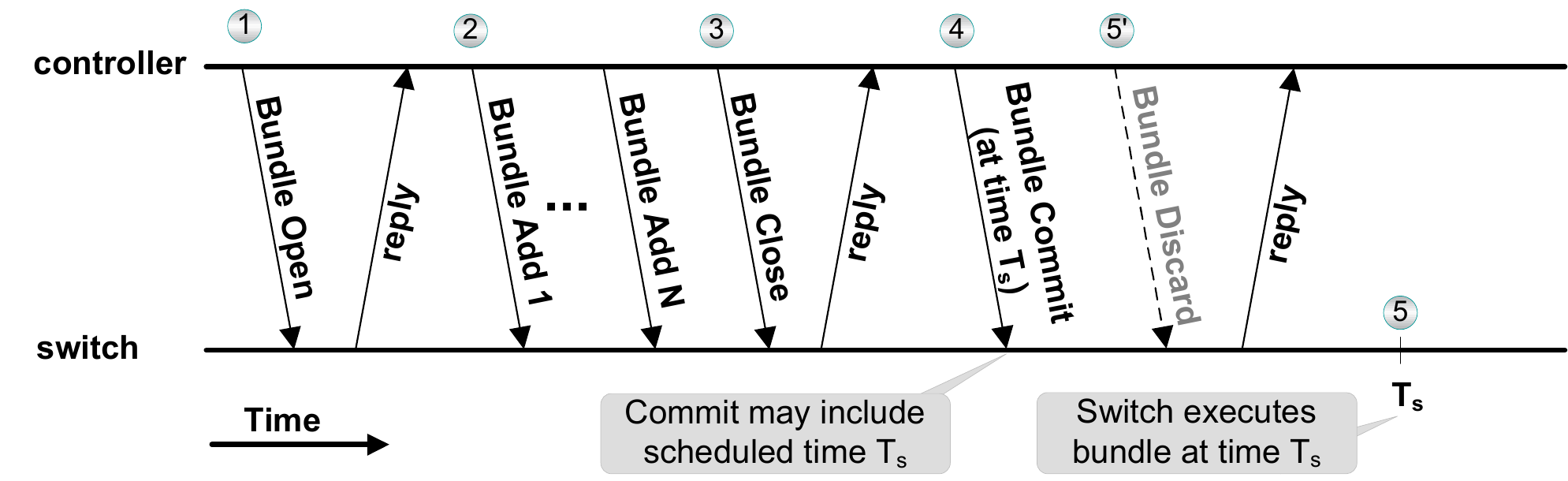}}
	\captionsetup{justification=raggedright}
  \caption{A \emph{Scheduled Bundle}: the \emph{Bundle Commit} message may include $T_s$, the scheduled time of execution. The controller can use a \emph{Bundle Discard} message to cancel the \emph{Scheduled Bundle} before time $T_s$.}
  \label{fig:Bundle}
\end{figure*}

\section{Scheduling Protocols}

One of the main goals of this project is to define extensions to standard network protocols, enabling practical implementations of the concepts we present. We defined a new feature in OpenFlow called \emph{Scheduled Bundles}~\cite{Time4Infocom,Time4Journal}, which enables time-triggered operations. We also defined a similar capability in NETCONF~\cite{OneClockNOMS}. These two features enable time-triggered operations in OpenFlow and in NETCONF, respectively. 

The two protocol extensions we defined can assist in translating the TimedSDN concepts from theory to practice, and open the door for future experimentation and potential deployment. Open source prototypes for these extensions~\cite{TimedSDNSource} are publicly available.

\subsection{OpenFlow Scheduled Bundles}

Our extension makes use of the OpenFlow~\cite{OpenFlow1.4} \textbf{Bundle} feature; a Bundle is a sequence of OpenFlow messages from the controller that is applied as a single operation. Our time extension defines \textbf{\emph{Scheduled Bundles}}, allowing all commands of a Bundle to come into effect at a pre-determined time. This is a generic means to extend all OpenFlow commands with the scheduling feature.

Using Bundle messages for implementing \timec\ has two significant advantages: (i)~It is a generic method to add the time extension to all OpenFlow commands without changing the format of all OpenFlow messages; only the format of Bundle messages is modified relative to the Bundle message format in~\cite{OpenFlow1.4}, optionally incorporating an execution time. (ii)~The Scheduled Bundle allows a relatively straightforward way to \emph{cancel} scheduled commands, as described below.

Fig.~\ref{fig:Bundle} illustrates the \emph{Scheduled Bundle} message procedure. In step 1, the controller sends a \emph{Bundle Open} message to the switch, followed by one or more Add messages (step 2). Every \emph{Add} message encapsulates an OpenFlow message, e.g., a \emph{FLOW\_MOD} message. A \emph{Bundle Close} is sent in step 3, followed by the \emph{Bundle Commit} (step 4), which optionally includes the scheduled time of execution, $T_s$. The switch then executes the desired command(s) at time $T_s$. 

As a result of our work, the capability to perform time-triggered updates has been incorporated into the OpenFlow 1.5 protocol~\cite{OpenFlow1.5}, and into the OpenFlow 1.3.x extension package~\cite{OpenFlow1.3ext}. Our time-enabled OpenFlow switch prototype~\cite{TimedSDNSource} was adopted by the ONF as the official prototype of Scheduled Bundles.\footnote{The ONF process for adding new features to OpenFlow requires every new feature to be prototyped.}

\subsection{The NETCONF Time Capability}

The NETCONF time capability is a generic extension of the NETCONF protocol that allows Remote Procedure Calls (RPCs) to include information about time. The NETCONF time capability has been published as an experimental IETF RFC~\cite{TimeCap}.

The time capability provides two main functions:
\begin{itemize}
	\item \textbf{Scheduling.} When a client sends an \verb|rpc| message to a server, the message may include the \verb|scheduled-time| parameter, indicating \emph{when} the RPC is expected to be executed. The server then starts to execute the RPC as close as possible to the scheduled time, and once completed the server can respond with an \verb|rpc-reply| message.
	\item \textbf{Reporting.} When a client sends an \verb|rpc| message to a server, the message may include a \verb|get-time| element, requesting the server to return the execution time of the RPC. In this case, after the server performs the RPC it responds with an rpc-reply that includes the \verb|execution-time| parameter, specifying the time at which the RPC was completed.
\end{itemize}

\section{Accurate scheduling methods}

One of the main challenges in the use of accurate time is to implement \emph{accurate execution} of events, i.e., guaranteeing that scheduled network updates are executed as close as possible to the time for which they were scheduled. Even if all the switches have perfectly synchronized clocks, executing events at their scheduled time may be challenging due to the nondeterministic nature of the switches' operating systems, and due to other running tasks.
Two accurate scheduling methods were defined and analyzed in this project, \timef\ and \onec.

\subsection{Accurate Scheduling using \timef}
In~\cite{Infocom-TimeFlip,TimeFlipToN} we introduced \timef s  as a way to accurately implement updates of hardware switches in a time-based manner, by encoding them 
in a Ternary Content Addressable Memory (TCAM). 

A TCAM is a memory that is commonly used in hardware switches. Every bit in the TCAM has three possible values, `0', `1', and `don't care'. The `don't care' value can be used for representing a range of values.

\begin{figure}[htbp]
  \centering
  \fbox{\includegraphics[width=.47\textwidth]{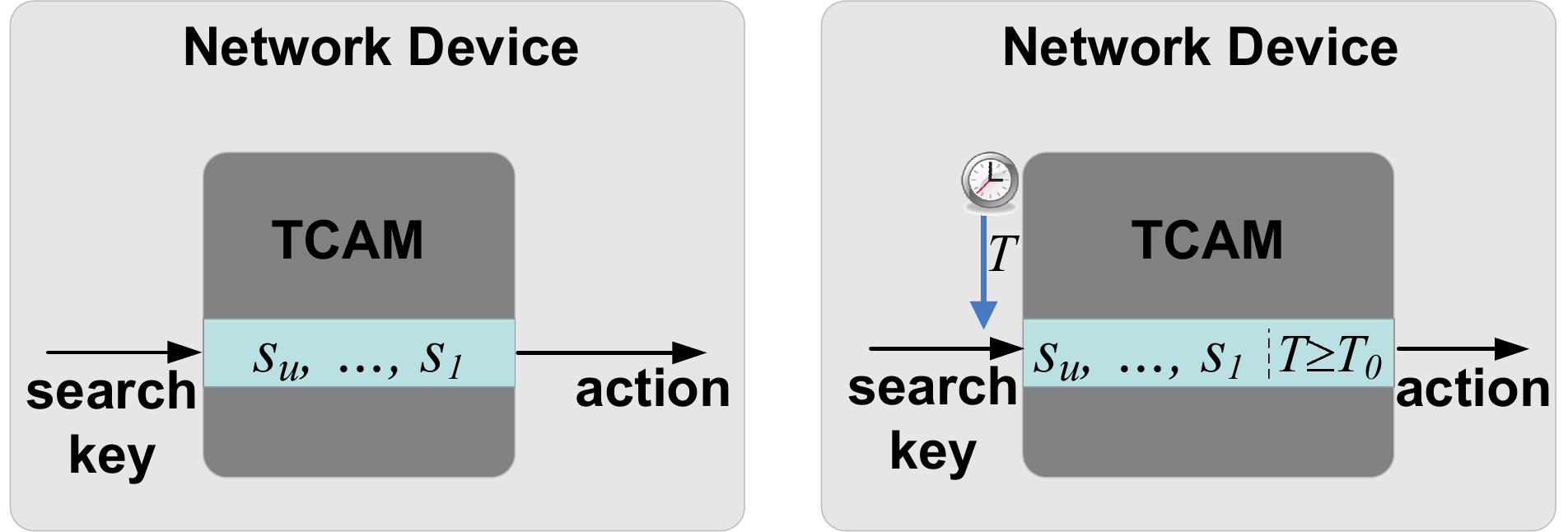}}
  \caption{TCAM lookup: conventional vs. \timef. \\ \timef\ uses a timestamp field, representing the time range $T\geq T_0$.}
  \label{fig:TimeTCAM}
\end{figure}

In this work we present a method that uses \proc{\textbf{TimeFlip}}s to perform \emph{accurate} \timeb\ network updates. We define a \timef\ to be a scheduled update that is implemented using TCAM ranges to represent the scheduled time of operation. We analyze TCAM lookups (Fig.~\ref{fig:TimeTCAM}) that take place in network devices, such as switches and routers. We assume that the device maintains a local clock, and that every packet that is received by the device is labeled with a timestamp recording its local arrival time.
Typically, TCAM search keys consist of fields from the packet header, as well as some additional metadata. In our setting, the metadata includes a timestamp~$T$. Hence, a TCAM entry can specify a range relative to the timestamp $T$, as a way of implementing \timeb\ decisions. 
The timestamp $T$ is not integrated into the packet, as it is only used internally in the device, and thus does not compromise the traffic bandwidth of the network device.

\timef s enable two important scenarios:

(i) \textbf{Network-wide coordinated updates.} If network devices use synchronized clocks, then \timef\ can be used for updating different devices at the same time,\footnote{Subject to the accuracy of the clock synchronization mechanism.} or for defining a set of scheduled updates coordinated in a specific order.

(i) \textbf{Atomic Bundle.} The \emph{Atomic Bundle} feature in OpenFlow~\cite{OpenFlow1.4} allows a set of configuration changes to be applied as a bundle, i.e., every packet should be processed either before any of the modifications have been applied, or after all have been applied. While implementing Atomic Bundles is typically non-trivial, \timef s allow a clean and natural way to implement Atomic Bundles; the set of configuration changes can be enabled at all times $T \geq T_0$ for some chosen time~$T_0$, and the timestamp $T$ defines when the bundle commands atomically come into effect.

	

\timef s require every TCAM entry to include a timestamp field. Moreover, representing a range of timestamp values may often require multiple TCAM entries to be used for each time range. One of the main contributions of this work is that we show that in practical conditions, a small number of timestamp bits are required to accurately perform a \timef\ using a small number of TCAM entries.

At the heart of our analysis lie two properties that are unique to \timeb\ TCAM ranges. First, by carefully choosing the scheduled update time, the range values can be selected to minimize the required TCAM resources. We refer to this flexibility as the \emph{scheduling tolerance}. Second, if there is a known bound on the installation time of the TCAM entries, then by using periodic time ranges, the expansion of the time range can be significantly reduced.

Our analysis provides a tight upper bound on the number of TCAM entries required for representing a \timef. We show that by correctly choosing the update time, the number of TCAM entries used for representing the timestamp range can be significantly reduced. We present an optimal scheduling algorithm; no other scheduling algorithm can produce a timestamp range that requires fewer TCAM entries.

\timef\ is a practical method of implementing accurate \timeb\ network updates
using \timeb\ TCAM ranges. \timef\ was tested in practice on a real-life device, a Marvell DX switch, showing that a \timef\ can be performed with a sub-microsecond accuracy, requiring very limited TCAM memory resources.

\subsection{Accurate Scheduling using OneClock}
OneClock~\cite{OneClockNOMS} is a prediction-based scheduling approach that uses timing information collected at runtime to accurately schedule future operations. OneClock allows a client to accurately schedule network operations without prior knowledge about the servers' performance. The approach is based on measuring the Elapsed Time of Execution (ETE) of each RPC, and using previous ETE measurements to predict the next ETE.

\begin{figure}[htbp]
	\ifdefined\CutSpace \vspace{-1mm} \fi
  \centering
  \fbox{\includegraphics[width=.2\textwidth]{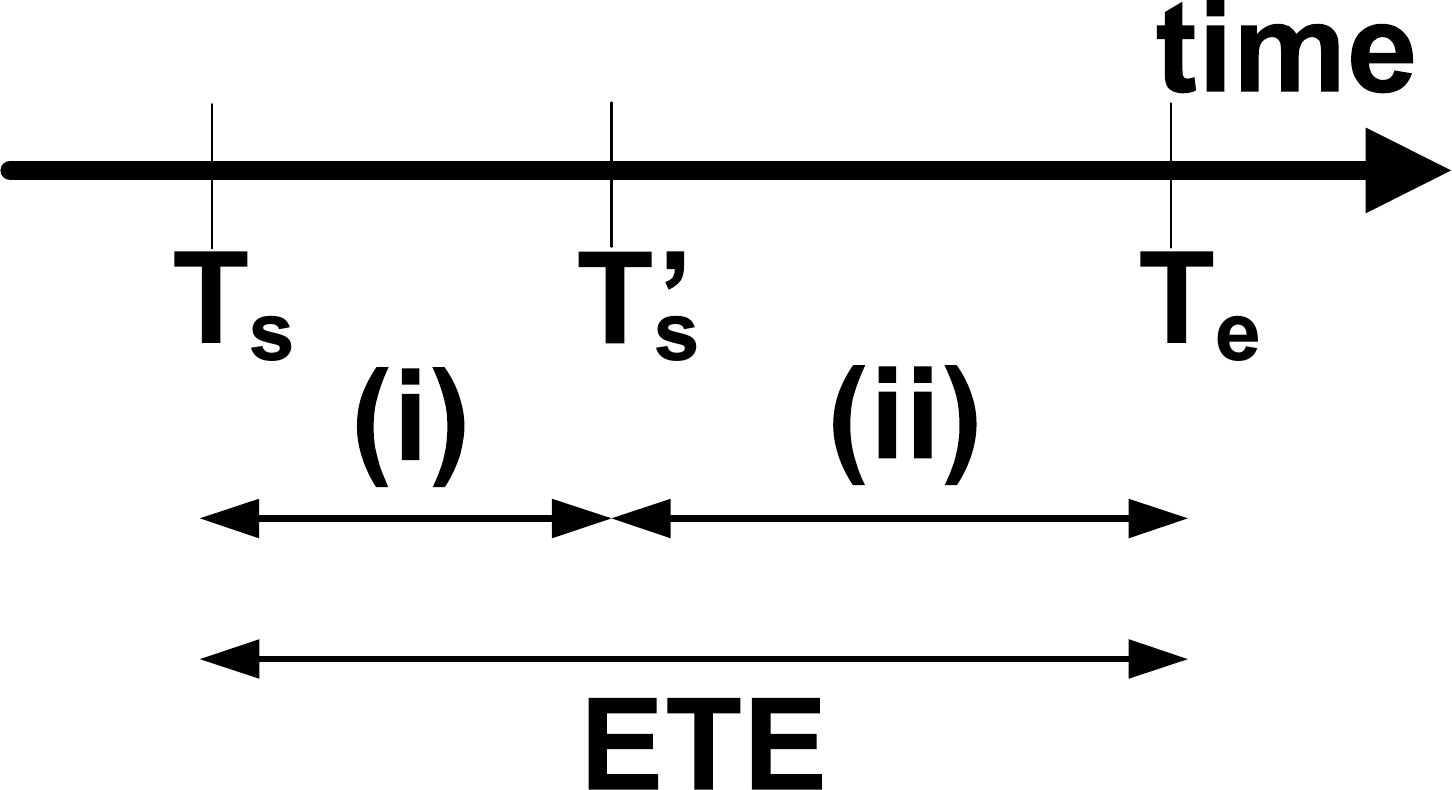}}
  \caption{Elapsed Time of Execution (ETE): ETE~=~${T_e-T_s}$.}
  \label{fig:ETE}
\end{figure}

\begin{figure}[htbp]
  \centering
  \fbox{\includegraphics[width=.47\textwidth]{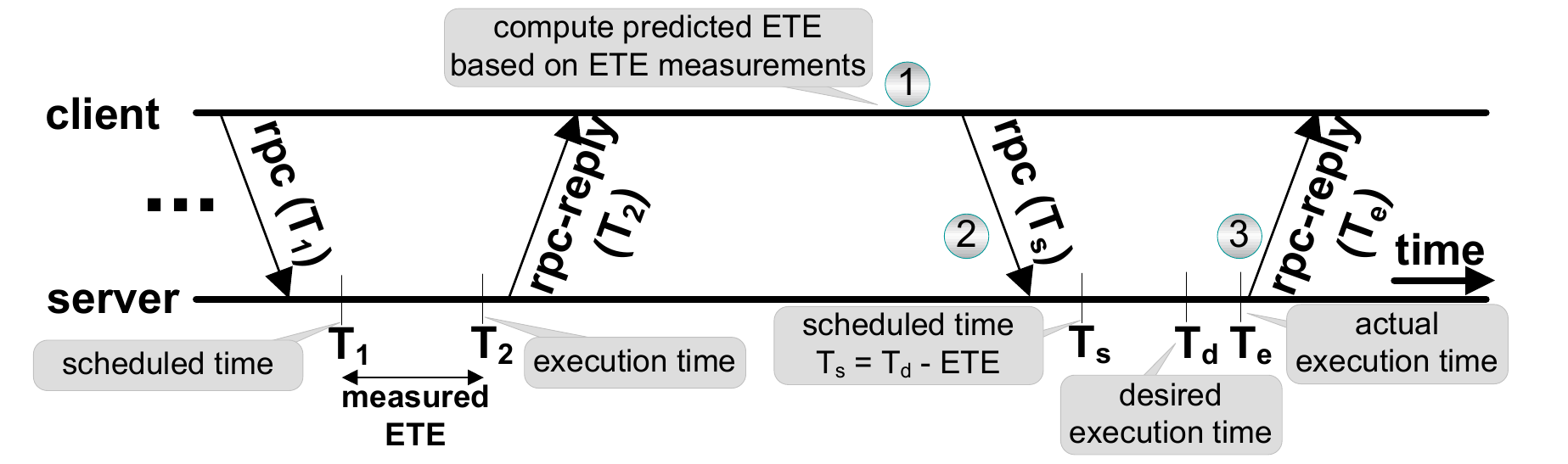}}
  \caption{Prediction-based scheduling: by predicting the ETE, a client can control when the RPC will be \emph{completed}.}
  \label{fig:Prediction}
\end{figure}

The ETE is defined to be $T_e-T_s$ (see Fig.~\ref{fig:Prediction}), where $T_s$ is the scheduled \emph{start time} of the RPC, and $T_e$ is the actual \emph{completion time} of the RPC. The \emph{actual} start time of the RPC is denoted by $T'_s$. Hence, as depicted in Fig.~\ref{fig:ETE}, the ETE is affected by two non-deterministic factors: (i) the server's ability to accurately start the operation, and (ii) the running time of the RPC. 

For each scheduled operation (see the numbered steps in Fig.~\ref{fig:Prediction}):\footnote{We follow the notation of~\cite{netconf}, where Remote Procedure Calls are denoted by uppercase RPC, and the messages that carry RPCs are denoted by lowercase \emph{rpc}.}

\begin{sloppypar}
\begin{enumerate}
	\item The client predicts the ETE of the next RPC based on previous measurements of the \emph{scheduled time} and \emph{execution time}. 
	\item For a given desired execution time, $T_d$, the client schedules the operation to be performed at~${T_d-ETE}$.
	\item The server reports the \emph{actual time of execution}, $T_e$, back to the client, allowing the client to use this feedback for scheduling future operations.
\end{enumerate}
\end{sloppypar}

Notably, OneClock allows a central entity to accurately schedule network operations in a heterogeneous environment, where the performance of the managed nodes is not necessarily known in advance.

Three prediction algorithms were analyzed in this work: (i)~an average-based algorithm, (ii)~a~fault-tolerant average (FT-Average), and (iii)~a~Kalman-Filter-based algorithm. In our experimental evaluation we found the simple FT-Average to be the most accurate algorithm in most of the experiments. The evaluation confirms that prediction-based scheduling provides a high degree of accuracy in diverse and heterogeneous environments, decreasing the prediction error by an order of magnitude compared to the na\"{\i}ve approach that does not use prediction.

\section{Clock Synchronization}
\label{SyncSec}

Clock synchronization is an essential piece in the puzzle. The concepts presented in the previous sections assume an underlying synchronization mechanism between the nodes in the network. Various synchronization mechanisms can be used, e.g., GPS-based synchronization, the Network Time Protocol (NTP)~\cite{mills2010rfc}, or the Precision Time Protocol (PTP)~\cite{IEEE1588}. Indeed, over the last few years PTP has become a mature and common technology for accurate synchronization.

The challenge of using a standard synchronization protocol such as PTP in an SDN environment lies in the fundamental difference between these two technologies. A key property of SDN is its centralized control plane, whereas PTP is a decentralized control protocol; a \emph{master clock} is elected by the Best Master Clock Algorithm (BMCA)~\cite{IEEE1588}, and each of the \emph{slaves} runs a complex clock servo algorithm that continuously computes the accurate time based on the protocol messages received from the master clock. Thus, if SDN switches function as PTP slaves, then in contrast to the SDN philosophy they are required to run complex algorithmic functionality, and to exchange control messages with other switches. Indeed, a hybrid~\cite{OpenFlow1.5} approach can be taken, where the SDN operates alongside traditional control-plane protocols such as PTP. Our approach is to adapt PTP to the SDN philosophy by shifting the core of its functionality to the controller.

\rptp~\cite{ispcsrptp,hotsdnrptp,rptpijnm} is a clock synchronization scheme that is adapted to the centralized SDN environment; in \rptp\ all nodes (switches) in the network distribute timing information to a single software-based central node (the SDN controller), that tracks the state of all the clocks in the network. For every switch $i$, the controller keeps track of $\offseti$, the offset between switch $i$'s clock and its local clock. Thus, if the controller needs to instruct switch~$i$ to perform an operation at time $T^c$ according to the controller's clock, it instructs the switch to perform the operation at time $T^c + \offseti$.

\begin{figure}[htbp]
  \centering
  \fbox{\includegraphics[width=.35\textwidth]{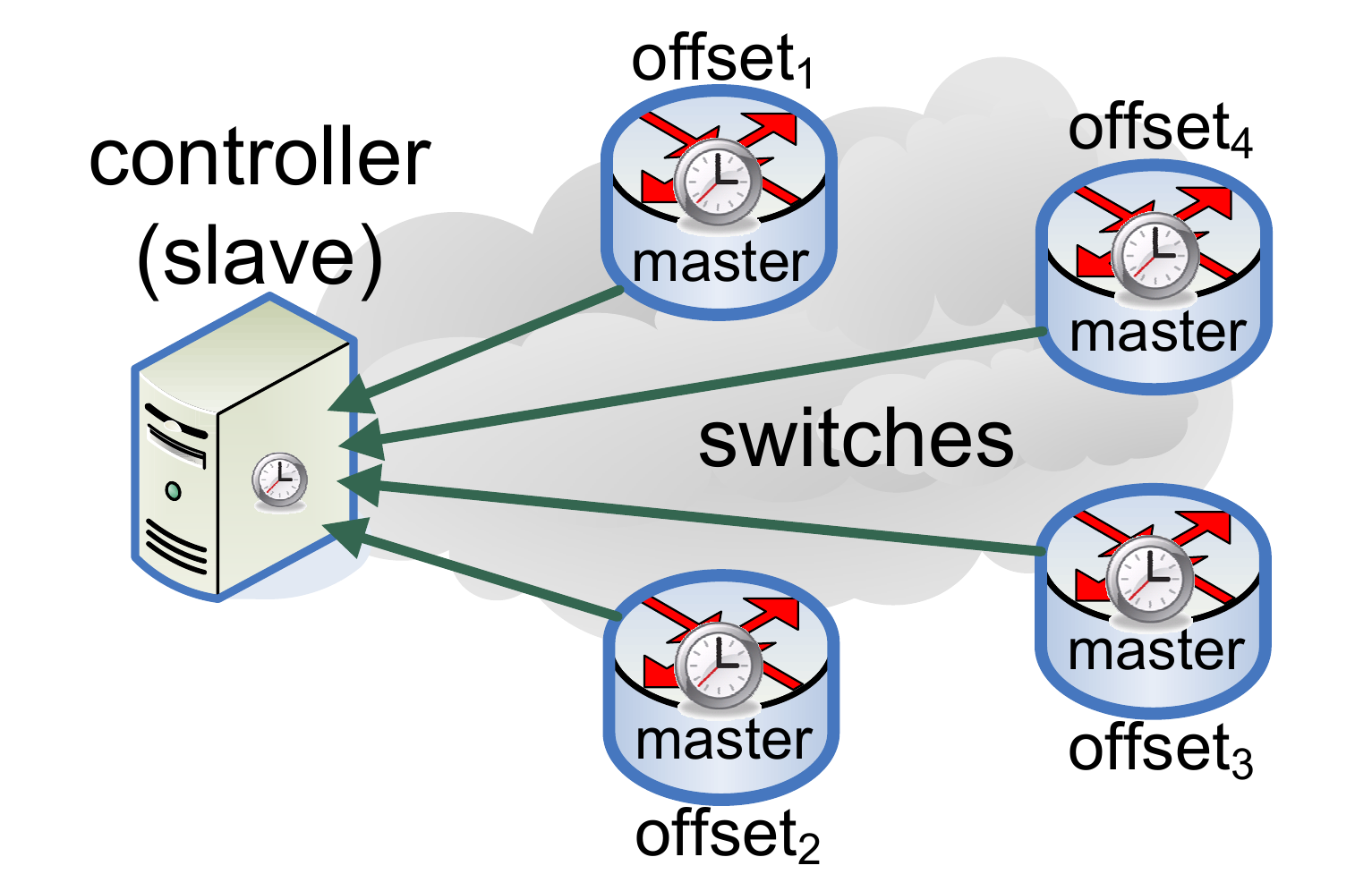}}
	\captionsetup{justification=raggedright}
  \caption{\rptp\ in SDN --- switches distribute their time to the controller. Switches' clocks are \emph{not} synchronized. For every switch $i$, the controller knows $\offseti$, the offset between switch $i$'s clock and its local clock.}
  \label{fig:rptp}
\end{figure}

In \rptp\ all computations and bookkeeping are performed by the central node, whereas the `dumb' switches are only required to periodically send their current time to the controller. In accordance with the SDN paradigm, the `brain' is implemented in software, making \rptp\ flexible and programmable from an SDN programmer's perspective.
Interestingly, \rptp\ can be defined as a PTP profile, i.e., a subset of the features of PTP. Consequently, \rptp\ can be implemented by existing PTP-enabled switches. An open source prototype of \rptp\ is publicly available~\cite{TimedSDNSource}.

\section{Conclusion}
This work investigated various aspects of using synchronized time in centrally-managed and software-defined environments. We believe there is great potential for future work, as clock synchronization technologies are continuously evolving, and as many centralized network applications may benefit from using time and synchronized clocks.

\bibliographystyle{ieeetr}
\bibliography{Timed}

\end{document}